# Assessing Cognitive Load on Web Search Tasks

Jacek Gwizdka, Dept. of Library & Information Science, School of Communication and Information, Rutgers University, New Brunswick, NJ, USA.
WebCog AT gwizdka.com

**Abstract:** Assessing cognitive load on web search is useful for characterizing search system features and search tasks with respect to their demands on the searcher's mental effort. It is also helpful for examining how individual differences among searchers (e.g. cognitive abilities) affect the search process. We examined cognitive load from the perspective of primary and secondary task performance. A controlled web search study was conducted with 48 participants. The primary task performance components were found to be significantly related to both the objective and the subjective task difficulty. However, the relationship between objective and subjective task difficulty and the secondary task performance measures was weaker than expected. The results indicate that the dual-task approach needs to be used with caution.

## Introduction and Background

Web search behavior is affected by the task, system, and individual searcher characteristics. Search tasks and their characterization have been a subject of recent systematic studies [1][2][3]. In particular, researchers have focused on the effects of task complexity and difficulty on information search process [4][5][6][7][8][9][10]. One kind of difficulty encountered by searchers is related to mental, or cognitive, requirements imposed by the search system or the task itself. Understanding factors that contribute to user's cognitive load on search tasks is crucial to identifying search system features and search tasks types that impose increased levels of load on users. As new interactive features are introduced into the information search systems we need to understand what determines their acceptance and why some evidently useful functions are not widely used. For example, user relevance feedback is a feature that has been reported to be avoided by users due to the heightened cognitive load [11]. Among other factors affecting search performance are the user's cognitive characteristics (e.g., [12][13]).

Methods used to date in assessing cognitive load included searcher observation, self-reports (e.g., using questionnaires, think-aloud protocols, and post-search interviews), dual-task techniques [14][15][16], and various approaches that employ external devices to collect additional data on users (e.g., eye-tracking, pressure-sensitive mouse and other physiological sensors [17]). The latter two groups of techniques have the advantage of enabling real-time, on-task data collection. However, the use of external devices can be expensive and impractical. Hence, the promise of dual-task (DT) method that allows for an indirect objective assessment of mental effort on the primary task. Only few studies employed this method to assess cognitive load in online search tasks (e.g., [14][15][8]). The article discusses the dual-task method as the technique for assessing cognitive load on web search tasks and presents research that contributes to better understanding of how objective task difficulty affects searchers' behavior and their perception of task difficulty.

## Research Objectives

We aim to understand cognitive load associated with performance of web search tasks. The current study examined dual-task method as an assessment technique of cognitive load on search tasks and considered the effects of selected individual differences and contribution of different types of "cognitive actions" (e.g., query formulation, search results inspection, reading individual web pages, relevance judgment) to the searcher's perception of task difficulty. This study extends our previous work [6][18] by including new variables into the examination of factors that affect subjective assessment of search task difficulty. In particular, the current study aimed to examine the following:

- relationships between the searcher's "cognitive activities" and subjective perception of task difficulty;
- which of the searcher's actions are good predictors of subjective task difficulty;
- whether performance on the search task is affected by the levels of task variables (e.g., objective difficulty);
- real-time assessment of cognitive load by employing dual task methodology;
- whether the selected cognitive abilities affect search task and dual task performance.

## Method

### Participants

Forty eight participants (17 females and 31 males; mean age 27 years) participated in question-driven, web-based information search study conducted in a controlled experimental setting. Participants were recruited from Rutgers University student population (undergraduate and graduate).

We assessed two cognitive abilities of the study participants, operation span (working memory performance) [19], and mental rotation (ability to manipulate mentally spatial images) [20]. The cognitive tasks (Table 1) were administered on a computer. These particular cognitive factors were selected as likely to affect searchers' performance on web tasks [21][22][23]. For the analysis, we split the values of cognitive task performance at median into high and low groups. Other individual factors included participant's age, gender, first language, and their Web search experience.

Table 1. Cognitive tasks used in the study to assess participants' abilities

| Cognitive Factor | Variable | Test Name | Short Description | Reference |
|---|---|---|---|---|
| **Working Memory - Operation Span** | **WM**: Operation Span ratio (0-1.0). Higher score→ higher ability | CogLab on CD (Wadsworth) | Operation Span is one of the measures of working memory performance. Operation span predicts verbal abilities and reading comprehension. | [19] |
| **Spatial Ability - Mental Rotation** | **SA**: A combined measure of Mental Rotation ratio of correct responses divided by Mental Rotation Mean Reaction Time (RT). Higher values→ higher ability | PsychExperiments (Dept. of Psychology, Mississippi University) | Mental Rotation is the ability to mentally manipulate spatial images. | [20][24][25] |

### User Tasks

The study search tasks were motivated by questions that described what information needed to be found and provided a context for the search. The tasks were designed to differ in terms of their difficulty and structure. Twelve questions were used in total, eight out of which were created by [9], while four simple fact-finding tasks were created for this study. Two types of search tasks were used: Fact Finding (FF) and Information Gathering (IG). The goal of a fact finding task is to find one or more specific pieces of information (e.g., name of a person or an organization, product information, a numerical value; a date). The goal of an information gathering task is to collect several pieces of information about a given topic. The tasks were also divided into three categories that depended on the structure of the underlying information need, 1) Simple (S), where the information need is satisfied by a single piece of information (by definition, simple task is of fact finding type); 2) Hierarchical (H), where the information need is satisfied by finding multiple characteristics of a single concept (a depth search); 3) Parallel (P), where the information need is satisfied by finding multiple concepts that exist at the same level in a conceptual hierarchy (a breadth search) [9]. By definition, there were five possible combinations of task types and structure: FF-S, FF-H, FF-P, IG-H, and IG-P.

Based on their characteristics, we categorized tasks into three levels of "objective" difficulty. FF-S was assigned low difficulty level, FF-P and FF-H middle-difficulty level, and IG-H and IG-P high difficulty level. We assigned three rather than five objective difficultly levels, because it is debatable whether the difference in the task structure between the parallel (P) and hierarchical (H) implies a difference in the task difficulty.

During the course of each study session, participant performed six tasks of differing type and structure (Table 2). For each task, participant was able to choose between two questions of the same type and structure but on different topics. We offered the choice to increase the likelihood of participants' interest in the question's topic. The order of tasks was partially balanced with respect to the objective task difficulty to obtain all possible combinations of low-medium-high and high-medium-low difficulty within the groups of three tasks (Table 2).

Table 2. Task rotations (for one rotation of search system).

| QR / Task Seq. | TSeq1 | TSeq2 | TSeq3 | TSeq4 | TSeq5 | TSeq6 |
|---|---|---|---|---|---|---|
| **QR1** | FF-S1 | FF-P1 | IG-H1 | FF-S2 | FF-H1 | IG-P1 |
| **QR2** | IG-H1 | FF-P1 | FF-S1 | IG-P1 | FF-H1 | FF-S2 |
| **QR3** | FF-S1 | FF-P1 | IG-H1 | IG-P1 | FF-H1 | FF-S2 |
| **QR4** | IG-H1 | FF-P1 | FF-S1 | FF-S2 | FF-H1 | IG-P1 |

A secondary task (DT) was introduced to obtain indirect objective measures of user's cognitive load on the primary search task [8]. A small pop-up window, controlled by a Java program written by us for this study, was displayed at a fixed location on a computer screen at random time intervals (15-29 seconds) and for a random period of time (5-9 seconds). The length of a cycle was thus between 20 and 38 seconds. The pop-up contained a word with a color name (Figure 1). The color of the word's font either matched or did not match the name of the color. Participants' were asked to click on the pop-up as soon as they noticed it. The click was performed either on the right (match between the color name and the font color) or on the left mouse button (no-match). The pop-up window disappeared after a random period of time or as soon as it was clicked on. Color names and font colors [26] were included in this task to ensure cognitive engagement of users and to avoid automaticity (perceptual and motor reaction to a visual stimulus). The secondary task involved motor action, as well as visuo-spatial and verbal/semantic processing. Similar types of processes were involved in performance on the primary search task. For example, processing a web page with search results or a page with an individual document that contains links involves understanding words (verbal/semantic), decision to click, moving mouse pointer (visuo-spatial, motor) and clicking on a desired link. The modalities of the primary task and the secondary task overlapped, and one could have reasonably assumed that different levels cognitive effort on the primary search task should be reflected in the differences of performance on the secondary task.

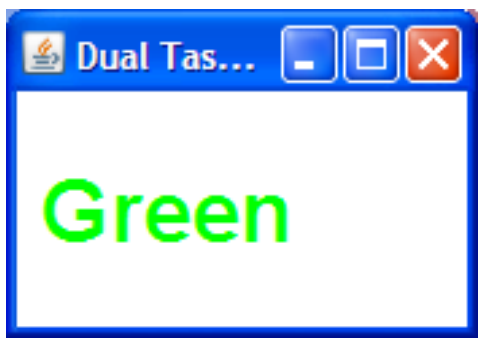

Figure 1. The secondary task pop-up window (not to scale).

## Search System

The search tasks were performed on the English Wikipedia. Two different search engines and interfaces were employed, U1: Google Wikipedia search, and U2: ALVIS Wikipedia search [27][28]. The search interface was switched after task 3. The four task rotations (Table 2) were repeated for two orders of user interfaces (U1/U2 and U2/U1). Thus there were a total of eight tasks and UI combinations.

## Procedure

Each study session was an hour and a half to two hours long and was conducted in a university lab on a personal desktop computer running Microsoft Windows XP operating system. Each session consisted of the following steps: introduction to the study, consent form, three cognitive tasks (cognitive style w-a, mental rotation and operation span), search task practice, secondary task practice, background questionnaire, six search tasks, and post-session questionnaire. Before and after each search task, participants answered a short set of questions about their familiarity with and interest in the subject area, about subjective perception of task difficulty (before and after), about their search satisfaction. Web pages that searchers considered relevant were bookmarked and tagged by them. User interaction with computer (the primary and the secondary task events, visited and bookmarked URLs, mouse and keyboard events, and screen cam) was recorded using Morae software and the secondary task program.

## Independent Factors

As presented above, the two main controlled factors were the objective task difficulty (OBJ_DIFF) and the search system (UI). The additional two independent factors were the levels of working memory (WM) and spatial ability (SA).

## Dependent Variables

**Behavioral measures (BE).** The recorded, time-stamped sequence of URLs was used to calculate measures of the searcher's behavior. In particular, of our interest were counts of visits to web pages related to "cognitive actions" (CA) such as, entering search queries, viewing search results and making decisions about what pages to read, reading web pages to assess their relevance to the task question, saving pages judged as relevant and entering tags to describe these pages. The measures based on web page visit counts were calculated for each search task and, with the exception of the total number of pages visited, did not include revisits to web pages. Revisits were accounted for by calculating two derived measures. 1) *Ratio of page revisits* [29] that was calculated using the ratio of unique pages to all pages visited in the following way: revisit_ratio = 1– uniq_nodes / total_nodes. The higher the revisit ratio, the more pages were revisited. Hence, the less efficient the searcher was. 2) *Navigation path linearity - stratum*. If we consider the individual web pages visited by searcher to be the nodes of a graph and the links actually followed by the searcher to be the graph edges, we can compute the graph properties, such as *stratum* [30]. Stratum was used to characterize searcher's behavior on web navigation tasks in past research studies ([31][32][33][34][13][6]). Stratum varies between zero and one. A value close to zero indicates a less linear navigation path; a value close to one indicates a nearly linear navigation path. We also calculated navigational speed as the average time spent on a web page. The above behavioral measures can be considered as belonging to two groups, Search Effort, and Search Efficiency [6]. The final behavioral measure was time on task (Table 3).

Table 3. Summary of behavioral variables (BE).

| Variable Group | Variable Name | Variable Description |
|---|---|---|
| **Search Effort** | tot_nodes | total number of web pages visited |
| | uniq_nodes | number of unique web pages visited |
| | resPg1_noRev | number of visits to first pages with search results (equal to the number of queries entered) |
| | resNext_noRev | number of visits to the subsequent result pages |
| | indRes_noRev | number of individual results visited |
| | bookmark_cnt | number of bookmarked individual result pages |
| | allCogActions | total of the four variables above (2*resPg1+resNext+indRes+bookmark_cnt) |
| **Search Efficiency** | revisit_ratio | ratio of revisits to web pages |
| | stratum | linearity of navigation path |
| | t_per_click | navigation speed: average time per web page |
| **Time** | duration | total time on each task |

**Secondary task performance (DT)**. We recorded searchers' interactions with the secondary task (DT). The following measures were derived: 1) Average reaction time to DT events; 2) Number of missed DT events; 3) Ratio of the total presence time of the DT pop-up window that was missed to the search task duration; 4) Ratio of all clicks on DT pop-ups to the number of all DT pop-ups; 5) Ratio of correctly clicked to all clicked DT pop-ups (dt_ratio_corr_to_clicks); 6) Ratio of clicks on DT pop-ups to the number of visited web pages during a search task (ratio_click_tot_nodes); and, a subjective measure, 7) Ratio of the estimated to the number of actual number of DT events (qa_ratio_click_count). These measures were expected to reflect cognitive load on the primary task.

**Subjective difficulty measure (SD)**. Upon the completion of all six search tasks, participants were asked to assess the difficulty of all search tasks by ranking the tasks on a 3 point difficulty scale (low-medium-high).

**Search Task Outcomes (TO)**. Three experts independently judged web pages that were bookmarked as relevant by the study participants. The experts assessed the *relevance* of the bookmarked documents and the extent to which a document covered answer to the question (*completeness*, also called *part of answer*). The inter-rater agreement assessed by employing Intra-class Correlation Coefficient was good to very good. For *relevance*, the average Intra-class Correlation Coefficient was 0.731 ($F(725,1450)=3.715$, $p<.001$). For *part of answer*, the average Intra-class Correlation Coefficient was 0.862 ($F(727,1454)=7.232$, $p<.001$).

## Expectations

We expected to find that:

- the objective and the searcher a posteriori assessed subjective task difficulty (SD) will be positively related;
- the subjective difficulty (SD), behavioral measures (BE) and performance on the secondary task (DT) should differ between the three levels of objective and subjective task difficulty.;
- better primary task outcomes will be associated with less difficult tasks;
- more difficult tasks (objective and perceived) will be associated with more searcher actions [6];
- performance on the "new" search system (ALVIS) will be worse (in terms of speed and task outcomes) then on the known system (Google);
- assuming that the observed behavior on the primary task reflects cognitive effort of a searcher, and that both primary and secondary tasks loaded on the same resources, then performance on the secondary task will be lower for more difficult primary tasks;
- performance on the primary and secondary task will be better (faster, more pages examined and more relevant results found) for higher levels of cognitive abilities.

## Results

### Objective and Subjective Task Difficulty

We first examined the relation between the objective and the subjective task difficulty. Their association was only medium strong (Spearman rho=.26, p<.001; Kendall's tau-b=.234, p<.001). The difference in the subjective difficulty levels among the three levels of objective task difficulty was significant (non-parametric Friedman test $\chi^2(2,N=94)=17$, $p<.001$). The relation was in the expected direction. However the differences between the mean values of subjective difficulty were smaller and skewed towards the low difficulty end of scale. Participants generally tended to underestimate task difficulty as compared to the objective difficulty created by search task design (Figure 2). Out of the total of 288 participant x task cases (48 participants times 6 tasks), 162 (57%) cases were rated as "low difficulty". This may reflect higher than expected Internet search experience among the study participants. 58% participants reported that they searched internet several times a day, while 21% reported that searched internet almost constantly. 94% agreed or strongly agreed that they were typically satisfied with their search results. It is a potential limitation of the study and we discuss it further in Conclusions.

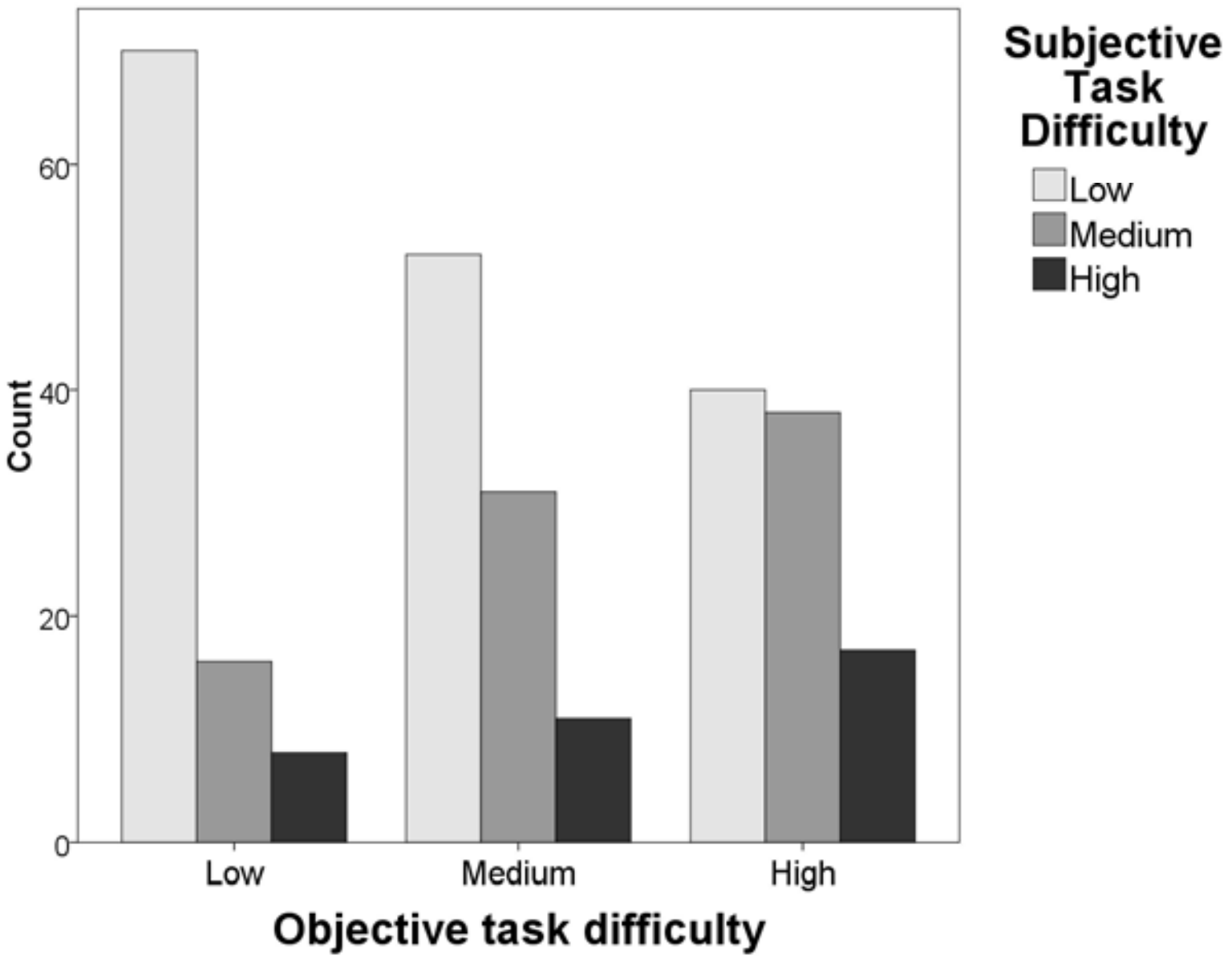

Figure 2. Objective and subjective task difficulty.

### Dependent Variables and Task Difficulty

We then examined[1] whether the means of the behavioral variables (BE), search task outcomes (TO), and secondary-task variables (DT) differed across the three levels of objective and subjective difficulty (Figure 3).

[1] One-way ANOVA was used for most variables, while non-parametric Kruskal-Wallis test was used for variables, whose distribution did not meet the criteria of analysis of variance (not normal, non-symmetrical).

**BE**: All individual behavioral variables, except the number of visits to the subsequent result pages, differed significantly for both objective and subjective task difficulty (selected statistics are presented Table 4). For example, the total number of cognitive actions was for low subjective difficulty tasks 13.8 actions less than for high difficulty tasks and 8 less than for medium difficulty (post-hoc Bonferroni test, p<.001), while for medium difficulty tasks it was 5.8 less than for high difficulty (p<.01).

**TO**: For search task outcomes, relevance differed significantly for both objective and subjective task difficulty, while completeness differed only for objective difficulty. The differences were in the expected direction, that is higher average relevance and more complete answers were achieved in less difficult tasks.

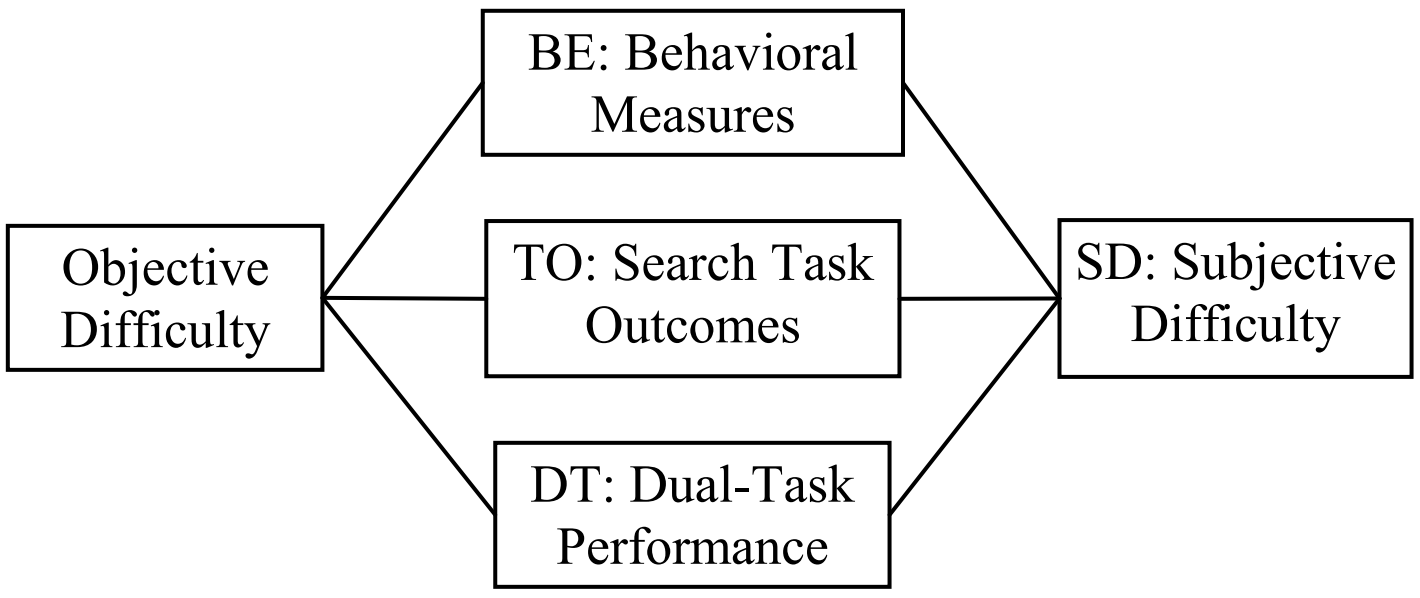

Figure 3. Examined differences in the means of dependent variables for objective and subjective task difficulty.

**DT**: We found that only the ratio of clicks on DT pop-up to the number of visited web pages (ratio_click_tot_nodes) differed significantly across the levels of objective task difficulty. The ratio for low difficulty tasks was 1.5, that is, the searchers clicked the secondary task pop-ups 50% times more than the number of web pages they visited in the primary task. This ratio for low difficulty tasks was 37% higher than for medium difficulty (post-hoc Bonferroni test, p<.05), while it was 22% higher than for high difficulty tasks (not a significant difference).

For the subjective task difficulty, two of the seven DT measures differed significantly. The ratio of correct to all DT clicks (dt_ratio_corr_to_clicks) was 94% for low and medium difficulty tasks, while for 87% for high difficulty tasks. For low difficulty tasks, the number of DT events tended to be overestimated (qa_ratio_click_count) by 30%, while for high and medium difficulty tasks the average number of estimated DT events was about right. The differences in the values of the significant DT measures between the levels of task difficulty were in the expected direction.

Table 4. Selected significant differences in variable values for objective and subjective task difficulty.

| Variable Group | Variable | for objective task difficulty | for subjective task difficulty |
|---|---|---|---|
| BE | all cognitive actions | $F(2,285)=41.7, p<.001$ | $F(2,280)=48.4, p<.001$ |
| | path linearity (stratum) | $F(2,285)=12.1, p<.001$ | $F(2,280)=9.5, p<.001$ |
| | navigation speed | $F(2,285)=11.1, p<.001$ | $F(2,280)=6.99, p=.001$ |
| | time on task (duration) | $F(2,285)=23.6, p<.001$ | $F(2,280)=42.3, p<.001$ |
| | unique web pages visited | $F(2,285)=25.2, p<.001$ | $F(2,280)=54.2, p<.001$ |
| | total web pages visited | $F(2,285)=22, p<.001$ | $F(2,280)=38, p<.001$ |
| TO | relevance | $\chi^2(2,N=277)=9.2, p<.01$ | $\chi^2(2,N=272)=18.3, p<.001$ |
| | completeness (part_of_answer) | $\chi^2(2,N=277)=10.1, p<.001$ | N/S |
| DT | correct clicks to all DT clicks | N/S | $\chi^2(2,N=281)=8.2, p<.05$ |
| | user estimate of DTs to actual DTs | N/S | $\chi^2(2,N=281)=8.3, p<.05$ |
| | clicks on DT to total pages | $\chi^2(2,N=288)=8.3, p<.05$ | N/S |

The relationships between BE, TO and DT variables and the task difficulties generally matched our expectations. However, the relationship between DT variables and objective and subjective difficulty was weaker than expected and only a couple of DT variables had a significant relationship.

We used linear regression to examine if BE, TO and DT variables can be used to predict objective task difficulty. Variables included in the model are shown Table 5. These variables were generally related to task outcomes, and included the number of marked relevant pages, the completeness of marked pages, and the average relevance. The model overall explains 28% of variance in the objective task difficulty. As expected, the more difficult the task, the more relevant pages were bookmarked and the less complete the result was. Surprising was the direction of relationship between relevance and objective task complexity. The higher the average relevance, the more difficult task.

Table 5. Predictors of objective task difficulty ($R^2$=.28).

| Variable Group | Variable | Stand. Beta coeff. | Incremental contrib. to variance explained |
|---|---|---|---|
| **BE** | Num. of bookmarks **** | 0.54 | +15% |
| **TO** | Completeness **** | -0.45 | +10% |
| | Relevance *** | 0.20 | +3% |

***p < .01 ****p < .001

To examine further whether measures of DT task performance are useful in predicting subjective difficulty on search tasks, we performed three regression analyses. Subjective task difficulty was the dependent variable, while independents (predictors) were as follows: 1) BE measures; 2) DT measures; 3) BE, TO and DT measures combined. $R^2$ for the obtained models was .3, .05, and .31 respectively. Thus, the combined model explains 31% of variance in the subjective task difficulty. Variables included in the model obtained with the combined set of predictors are shown Table 6. Clearly, BE measures are much stronger predictors of subjective task difficulty than DT measures.

Table 6. Combined regression analysis. Predictors of subjective task difficulty ($R^2$=.31).

| Variable Group | Variable | Stand. Beta coeff. | Incremental contrib. to variance explained |
|---|---|---|---|
| **BE** | Num. of individual results examined **** | 0.50 | +23% |
| | Num. of first search result pages examined **** | 0.21 | +3% |
| **DT** | Ratio of correct to all clicks on DT pop-up *** | -0.14 | +2% |
| **BE** | Num. of bookmarks ** | -0.15 | +2% |

**p < .05 ***p < .01 ****p < .001

## Effects of the Independent Factors

Independent factors included two factors that were controlled in the study (objective task difficulty and search system), as well as two factors that characterized cognitive abilities (working memory and spatial ability). To examine the main effects[2] and the interaction effects of these factors on dependent variables (BE, TO, DT, and SD), we performed a series of Unianova analysis with these four factors and one dependent variable at a time (
Figure **4**).

Table 7 summarizes the results of these analyses. Detailed statistics are included in the Appendix in Table 9. Table 7 also includes relationships between dependent variables and the subjective task difficulty that were presented in the previous section.

[2] Our use of "effect" and "affected" in presentation of the relationships among the independent factors and dependent variables does not imply their causal relationship. However, the underlying conceptual model lets us infer possible causal relationships between independent factors and dependent factors, and between independent factors, searcher behavior and subjective difficulty.

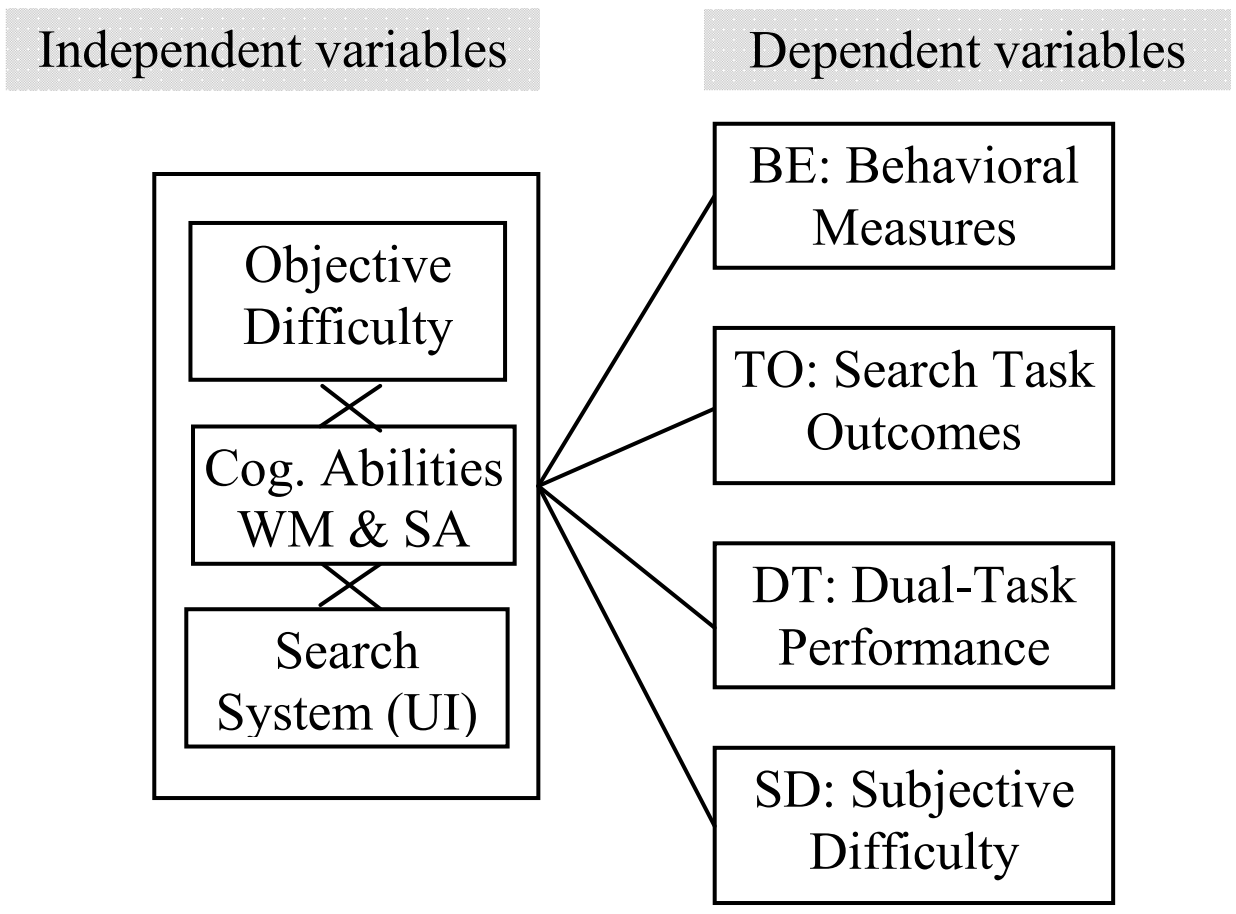

Figure 4. Models of examined relationships.

Table 7. Summary of the factors effects (main effects and UI & Cognitive Abilities interaction effects). Values in the cells are partial eta squared expressed as percentages ($\eta_p^2$ %).

| Variable Group | Variable | Obj. Difficulty | Subj. Difficulty | UI | WM | SA | WM x SA | UI x WM |
|---|---|---|---|---|---|---|---|---|
| **BE: effort** | all cognitive actions | 20 | 25.4 | | 1.7 | | | |
| | uniq pages | 17 | 27 | | 2.4 | | | 2 |
| | tot pages | 14.7 | 24 | | | | | 2.4 |
| | queries / result lists (unique) | 12.4 | 16 | | | | | |
| | individual results | 15.2 | 26 | | | | | |
| | saved relevant bookmarks | 18.4 | 5 | | | | | 2.7 |
| **BE: efficiency** | navigation speed | 6 | 5.2 | 1.5 | | | | |
| | linearity of navigation path | 9 | 9.7 | 6.3 | | | | 3.5 |
| | revisits to web pages | 4.2 | 6.3 | 10 | | | | 2.2 |
| **BE: time** | time on task (duration) | 13 | 20.4 | | 2 | | | |
| **TO** | relevance | 4.4 | 2.1 | | | | 2 | |
| | completeness (part of answer) | 5 | | | | | | |
| **Subj. Diff.** | subjective task difficulty | 6.3 | N/A | 4.2 | 1.7 | | | |
| **DT: general performance** | user clicks to all DT events | | | | 3 | 3 | 5 | |
| | average RT | | | | 5 | 2 | 3 | |
| | missed DTs / clicks | | | | 3 | 3 | 6 | |
| | correct clicks to all DT clicks | | * | | 5 | 4 | 8 | |
| **DT: relative** | clicks on DT to total pages | 2.4 | | | | | | |
| **DT: subjective** | user estimate of DTs to actual DTs | | * | | | | | |

All effects were significant at p<.05 or better.

We observe that the search task outcomes (TO) were affected by objective task difficulty (OBJ_DIFF) and by interaction between WM and SA. We saw the former relationship in the regression results presented in the previous section. The latter relationship was unexpected; for low spatial ability the average relevance was higher for low working memory and lower for high working memory. We defer discussion of cognitive ability effects until after all relationships are presented. All behavioral measures (BE) were affected by objective difficulty, while some were also affected by working memory. Behavioral measures related to search efficiency were additionally affected by the search system and its user interface. As described in the previous section, subjective difficulty (SD) was affected by objective difficulty, but it was also affected by the search system and its user interface and by working memory. Most dual-task measures (DT), except the relative measure (ratio_click_tot_nodes) and the subjective assessment (qa_ratio_click_ount), were affected only by cognitive abilities (both working memory, spatial ability and their interaction). Two of DT variables, ratio of correctly clicked to all clicked DT pop-ups (dt_ratio_corr_to_clicks) and user estimated DT count to actual (qa_ratio_click_count), were also related to the subjective difficulty. While the relative measure of clicks on DT pop-ups to the number of visited web pages (ratio_click_tot_nodes) was affected only by objective difficulty.

The direction of the main effects of UI and OBJ_DIFF was as expected, with an exception of navigation speed. Participants were slower in the low difficulty tasks. The effects of cognitive abilities and the pattern of interaction effects and their direction were more complex. Most unexpected relationships seemed to be related to working memory effects. Participants with higher levels of WM tended to spend more time on primary task, visit more pages, and perform more cognitive actions. In particular, high WM participants visited more pages when using Alvis than when using Google. This higher number of examined pages resulted in more saved bookmarks in Alvis than in Google, where the WM level did not make significant difference. Higher working memory ability was also generally associated with lower levels of DT performance. For example, for searchers high on SA, WM did not make any difference in the average reaction time to secondary task. In contrast, for low SA, WM differentiated between the slower (high WM) and faster people (low WM). High WM participants also perceived tasks as more difficult.

Effect sizes (strength of associations) were estimated by calculating partial eta squared[3] (Table 7). The strongest association was between subjective difficulty and behavioral measures that express search effort (15%-25%), while the association between objective difficulty and these measures was somewhat less strong (12%-20%). Interestingly, this relationship between subjective difficulty and objective difficulty and behavioral variables was reversed for the number of bookmarked relevant pages, with which the objective difficulty was more strongly associated than subjective difficulty (18.4% vs. 5%). The strongest association between the search system (UI) and search efficiency was for the ratio of revisits, where it was higher than the association of objective difficulty (10% vs. 4.2%). For path linearity, the relationship was reversed, stronger for objective difficulty (9%) and weaker for UI (6.3%). The association among cognitive abilities and behavioral variables was generally weak (2-3%).

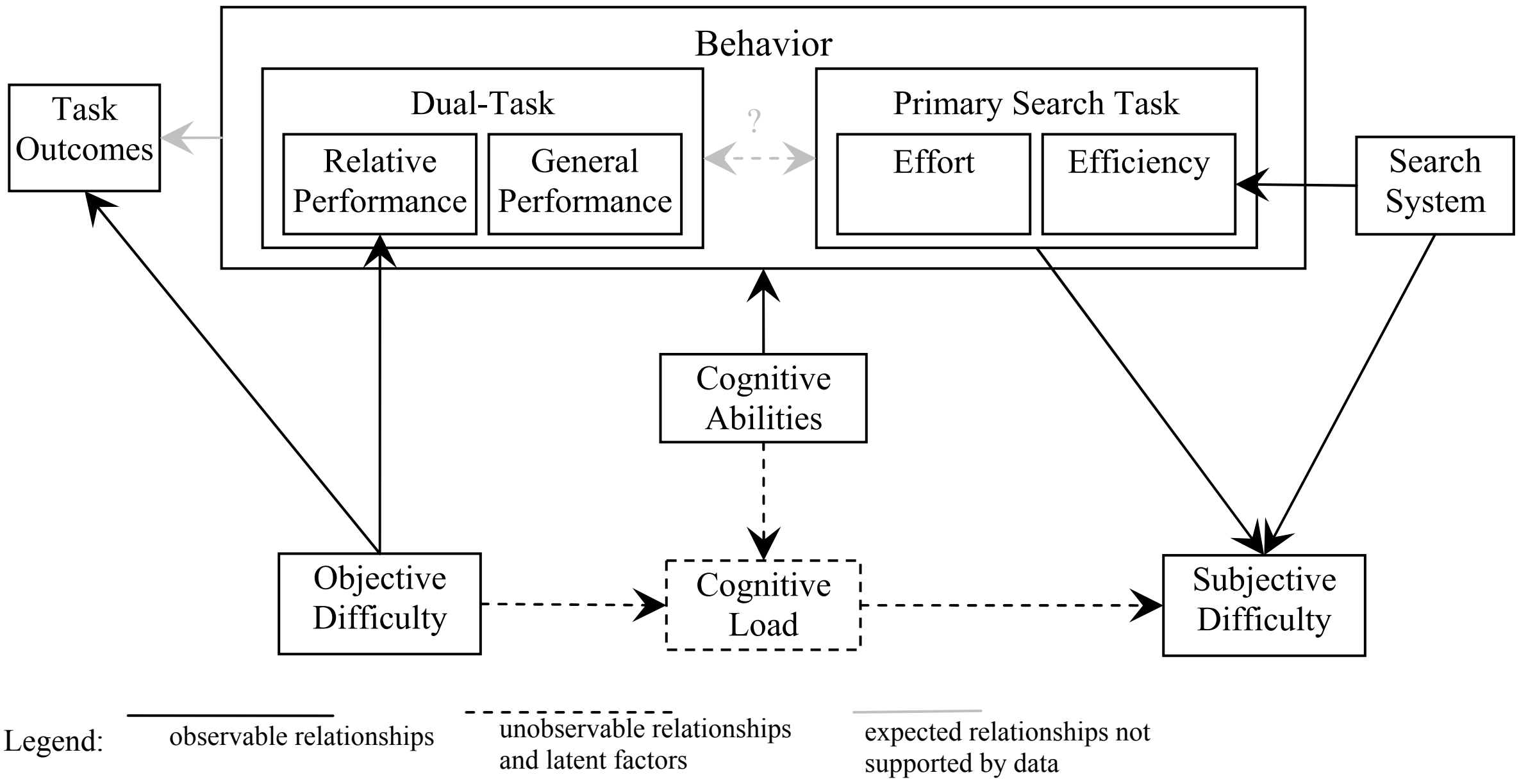

Figure 5. Observed and anticipated relationships.

## Conclusions and Discussion

Research presented in this article examined relationships among objective task difficulty, searcher behavior ("cognitive actions"), two cognitive abilities, and subjective task difficulty (Figure 5). We also examined dual-task method as an assessment technique of cognitive load on web search tasks.

As expected, subjective task difficulty was related to objective difficulty, however, the association was only medium strong and participants had a tendency to underestimate task difficulty. This tendency can be plausibly explained by noticing the high internet search experience among the participants and their relatively young age. Both objective and subjective difficulty were strongly associated with the searchers' behavior (time on task, cognitive actions, and the searchers' efficiency in their performance). In the case of objective difficulty, this was an expected effect of the designed task variability on the search effort and on efficiency of task performance. In the case of subjective difficulty, it was the effect of the searcher's effort on their perception of task difficulty. Search task outcomes were affected by the objective task difficulty, and were only weakly related to the subjective task perception. Task outcomes (relevance and completeness) are considered objective[4], and were not known to the searchers at the time of task completion. This relationship confirms the expected effect of task variables on outcomes. Subjective task difficulty was more strongly related with user effort than objective difficulty. This may indicate that the subjective task difficulty reflects more truly the searcher's effort (e.g. cognitive effort) than the objective difficulty and that some effort is not influenced by the latter. The additional variance in effort that was reflected in the subjective difficulty may be due to individual differences among users. This suggests that the subjectively perceived difficulty may be more strongly associated with unobservable cognitive load than the objective difficulty.

[3] Partial eta squared needs to be interpreted with caution as the individual components are not additive and the total may be greater than 1.0. We use $\eta_p^2$ as an indicator of relative differences among variables in their strength of association.

The search systems and the associated user interfaces affected primarily the search efficiency. This could be plausibly explained by the effect of learning of a new system (ALVIS) by participants. This explanation seems to be confirmed by the effects of working memory.

The modality of the primary task and the secondary task were designed to be similar to each other (they loaded on the visuo-spatial and verbal-semantic subsystems). One could thus reasonably expect an interference between the tasks and, as a result of the central capacity limitations, a drop in performance on the secondary task with increased load on the primary task (search task). accordingly, we expected to find a significant relationship between the secondary task measures (DT), objective and subjective task difficulty. A virtual absence of such significant relationships is likely due to the relatively low difficulty of the search tasks (approximately 57% of task instances were rated as low difficulty). Usefulness of the secondary task performance in the assessment of cognitive load depends on task load. In conditions when the total load of primary and secondary tasks is not sufficiently high, the DT measures may not work as well as expected. Most of the examined dual task measures, including the "standard" reaction time, were found to be affected not by the task difficulty, but only by the cognitive abilities. However, a couple of other DT measures were related to the subjective task difficulty of the primary task. A secondary task outcome measure (ratio of correct to all clicks) and a subjective measure (ratio of user estimated secondary task events to their actual number) were found to be associated with the subjective difficulty. A more interesting relationship was found between objective task difficulty and a relative measure of clicks on the secondary task to the number of visited web pages. This measure reflects the searcher's performance on the secondary task in relation to their performance (in the sense of effort) on the primary task. This type of measure seems to be promising in assessing objective load on tasks. Overall, these results indicate that the DT technique needs to be used with caution and that secondary task measures should be carefully constructed and tested.

Although care was taken to vary the search task difficulty, the tasks may have been overall too easy for the study population – college students, who are almost constantly online. Additionally, this population may be used to dealing with various issues in web search. That kind of experience could have skewed their assessment of web search task difficulty. Subsequent experiments should employ more difficult search tasks, possibly in combination with different user populations, examine other dual-task measures, and, in particular, other measures that capture the relative performance on the secondary and primary tasks.

The analysis presented in this paper was performed at the search task granularity. Future work should examine how the cognitive load changes between the different stages of web search tasks (e.g, query formulation, search result list examination, content reading) and how these changes are affected by the searcher's cognitive abilities and styles.

## Acknowledgements

This research was sponsored, in part, by a grant from Rutgers University Research Council #RCG202130.

[4] They are objective, insofar as the expert judgments can be relied upon. As reported, the agreement among the judges was good to very good.

## Appendix.

Table 8. Selected search tasks used in the study (one for each combination of task type and structure).

| Type | Question text |
|---|---|
| FF-S | You love history and, in particular, you are interested in the Teutonic Order (Teutonic Knights). You have read about their period of power, and now you want to learn more about their decline. What year was the Order defeated in a battle by a Polish-Lithuanian army? |
| FF-H | A friend has just sent an email from an Internet café in the southern USA where she is on a hiking trip. She tells you that she has just stepped into an anthill of small red ants and has a large number of painful bites on her leg. She wants to know what species of ants they are likely to be, how dangerous they are and what she can do about the bites. What will you tell her? |
| FF-P | As a history buff, you have heard of the quiet revolution, the peaceful revolution and the velvet revolution. For a skill-testing question to win an iPod you have been asked how they differ from the April 19th revolution. |
| IG-H | You recently heard about the book "Fast Food Nation," and it has really influenced the way you think about your diet. You note in particular the amount and types of food additives contained in the things that you eat every day. Now you want to understand which food additives pose a risk to your physical health, and are likely to be listed on grocery store labels. |
| IG-P | Friends are planning to build a new house and have heard that using solar energy panels for heating can save a lot of money. Since they do not know anything about home heating and the issues involved, they have asked for your help. You are uncertain as well, and do some research to identify some issues that need to be considered in deciding between more conventional methods of home heating and solar panels. |

Table 9. Selected statistics from the analysis of independent factors effects (main effects and UI & Cognitive Abilities interaction effects).

| Variable Group | Variable | OBJ_DIFF | UI | WM | SA | WM x SA | UI x WM |
|---|---|---|---|---|---|---|---|
| BE: effort | all cognitive actions | F(1,269)=32.6, p<.001 | | F(1, 269)=32.6, p<.001 | | | |
| | uniq pages | F(1, 269)=26.7, p<.001 | | F(1, 269)=6.5, p<.01 | | | F(1, 269)=5.3, p<.05 |
| | tot pages | F(1,269)=23.2, p<.001 | | | | | F(1, 269)=6.5, p<.05 |
| | queries / result lists (unique) | F(1,269)=19.1, p<.001 | | | | | |
| | individual results | F(1,269)=24, p<.001 | | | | | |
| | saved relevant bookmarks | F(1, 269)=30.3, p<.001 | | | | | F(1, 269.6, p<.01 |
| BE: efficiency | navigation speed | F(1,269)=8.2, p<.001 | F(1, 269)=4.2, p<.05 | | | | |
| | linearity of navigation path | F(1,269)=13.1, p<.001 | F(1, 269)=18.1, p<.001 | | | | F(1, 269)=9.9, p<.001 |
| | revisits to web pages | F(1,269)=5.8, p<.01 | F(1, 269)=29.8, p<.001 | | | | F(1, 269)=6, p<.05 |
| BE: time | duration | F(1,269)=19.8, p<.001 | | F(1, 269)=7.7, p<.01 | | | |
| TO | relevance | F(1,261)=6, p<.01 | | | | F(1, 261)=4.5, p<.05 | |
| | completeness | F(1, 261)=6.8, p=.001 | | | | | |
| Subj. Diff. | subjective task difficulty | F(2,264)=8.9, p<.001 | F(1, 264)=11.4, p=.001 | F(1, 264)=4.6, p<.05 | | | |
| DT: general performance | user clicks / all DT events | | | F(1,269)=8.6, p<.01 | F(1,269)=9.1, p<.01 | F(1,269)=16.6, p<.001 | |
| | average RT | | | F(1,262)=7.6, p<.01 | F(1,262)=4.4, p=.05 | F(1, 262)=7.2, p<.01 | |
| | missed DTs / clicks | | | F(1,269)=8.6, p<.01 | F(1,269)=9.1, p<.01 | F(1,269)=16.6, p<.001 | |
| | correct clicks to all DT clicks | | | F(1,262)=13.3, p<.001 | F(1,262)=10.6, p=.001 | F(1,262)=21.7, p<.001 | |
| DT: relative | clicks on DT to total pages | F(1,269)=3.3, p<.05 | | | | | |